\documentclass[fleqn,10pt]{wlscirep}
\usepackage[utf8]{inputenc}
\usepackage[T1]{fontenc}
\usepackage{bm}
\title{Anisotropic magnetoresistance and magnetic field-tunable Weyl nodes in Weyl metal SrRuO$_{3}$ thin films}

\author[1,$\dag$]{Uddipta Kar}%\thanks{These authors contributed equally to the work.}
\author[1,$\dag$]{Akhilesh Kr. Singh}%thanks{These authors contributed equally to the work.}
\author[1,$\dag$]{Elisha Cho-Hao Lu}%\thanks{These authors contributed equally to the work.}
\author[2]{P. V. Sreenivasa Reddy}
\author[1]{Fu-En Cheng}
\author[1]{Wazid Ahmed}
\author[3]{Song Yang}
\author[3]{Chun-Yen Lin}
\author[3]{Chia-Hung Hsu}
\author[2,4,*]{Guang-Yu Guo}
\author[1,*]{Wei-Li Lee}

\affil[1]{Institute of Physics, Academia Sinica, Nankang, Taipei 11529, Taiwan}
\affil[2]{Department of Physics, National Taiwan University, Taipei 10617, Taiwan}
\affil[3]{Scientific Research Division, National Synchrotron Radiation Research Center, Hsinchu 30076, Taiwan}
\affil[4]{Physics Division, National Center for Theoretical Sciences, Taipei 10617, Taiwan}
\affil[$\dag$]{These authors contributed equally to the work}
\affil[*]{e-mail: wlee@phys.sinica.edu.tw; gyguo@phys.ntu.edu.tw}

\begin{abstract}
Weyl semimetals are a unique class of topological materials, possessing Fermi-arc surface states and exhibiting the chiral anomaly effect. The chiral anomaly refers to non-equilibrium charge transfer within a Weyl-node pair of opposite chirality under the condition of aligned electric and magnetic fields ($\bf{E} \parallel \bf{B}$), leading to non-conserved chiral charges and thus enhanced electrical conductivity. In experiments, such an enhanced conductivity due to the chiral anomaly manifests as a negative longitudinal magnetoresistance (MR) when the external field $\bf{H}$ is applied along the bias current direction $\bf{I}$. In this work, we present rigorous $\phi$- and $\alpha$-dependent magnetotransport measurements to investigate such a negative longitudinal MR due to the chiral anomaly in a sunbeam-shaped device fabricated from an untwinned Weyl metal SrRuO$_{3}$ (SRO) thin film. Here, $\phi$($\alpha$) represents the angle between $\bf{I}$ and the in-plane $\bf{H}$(SRO monoclinic [001]$_{\rm o}$). Unusual $\phi$ dependences of in-plane MR and Hall effects were uncovered at low temperatures, accompanied by the emergence of the fourfold-symmetric component in the in-plane MR. These results indicate that the chiral anomaly and resistivity anisotropy in SRO play important roles. In particular, the dramatic variation of Weyl nodes near the Fermi level through magnetic field manipulation of the magnetization orientation, as revealed by band structure calculations, is consistent with the observed in-plane MR and Hall effect.
\end{abstract}
\begin{document}

\flushbottom
\maketitle

\thispagestyle{empty}

\section*{Introduction}
Magnetic topological materials with broken time-reversal symmetry are known to exhibit a number of remarkable physical effects, such as the quantum anomalous Hall effect in a magnetic topological insulator \cite{Chang2013,Kou2014,Checkelsky2014} and the chiral anomaly in a magnetic Weyl semimetal (WSM) \cite{Armitage2018,Hirschberger2016,BFB2022}. The internal field from magnetization ($\bf{M}$) results in the formation of Dirac mass domain walls on the surface states of a magnetic topological insulator, giving rise to unique, dissipationless chiral edge modes in a zero external field \cite{Chu2011,Yasuda2017}. On the other hand, the lifting of spin subband degeneracy in a magnetic WSM results in nontrivial band crossing points, i.e., Weyl nodes, which can be sensitive to the magnitude and the direction of the system's $\bf{M}$. Under the condition of $\bf{E} \parallel \bf{H}$ with a long intervalley lifetime, the $n$ = 0 chiral Landau levels for a Weyl-node pair disperse along $\bf{H}$ and result in a nonzero total chiral charge \cite{SS2013,Burkov2015}, which acts as an additional charge source and gives rise to an enhanced conductivity with a magnitude inversely proportional to the energy difference between the Weyl nodes and the Fermi surface. In addition, several intriguing magnetotransport effects associated with the unique Fermi-arc surface states have been reported in various magnetic WSM systems \cite{Kaneta2022,Kar2023,Matsuki2024,Xu2025}. Consequently, a magnetic WSM can be an ideal platform to realize the intriguing tuning of the electronic band structure and thus the Weyl-node distribution near the Fermi surface simply via the control of the $\bf{M}$ orientation, which is subsequently accompanied by a large variation in the associated magnetoconductivity response. Furthermore, the onset of a magnetic transition and magnetism typically results in a lower-symmetry state, thereby removing the symmetry constraints for certain physical effects. For example, it has been shown that mirror plane symmetry can be broken by an internal field from $\bf{M}$ \cite{Torre2021}, where the nonlinear Hall effect due to the Berry curvature dipole may then emerge without the constraints imposed by mirror plane symmetry \cite{Sodemann2015,Ma2019,KarPRX2024}. 

Historically, the enhancement of conductivity under the $\bf{H} \parallel \bf{I}$ condition due to the chiral anomaly has been demonstrated in several topological systems with conical bands \cite{Huang2015,Xiong2015}. However, rigorous $\bf{I}$-orientation-dependent magnetoconductivity measurements remain lacking, as they require the growth of highly crystalline and single-structural domain topological WSMs in thin-film form. Here, we report comprehensive magnetoconductivity measurements in a sunbeam-shaped device fabricated from untwinned and monoclinic thin films of the ferromagnetic Weyl metal SrRuO$_{3}$ (SRO) with an exceptionally high residual resistivity ratio (RRR) and low residual resistivity (RR). The monoclinic structure of the SRO thin film and the corresponding pseudocubic unit cell are shown by the black dashed lines and light-blue solid lines, respectively, in Fig. \ref{OFPMR}(a). Figure \ref{OFPMR}(b) shows an optical image of a fabricated sunbeam-shaped SRO device, where the patterned SRO Hall-bar devices are enclosed by the black dashed lines shown in the inset blowup view. The chiral anomaly and the in-plane Hall effect can then be rigorously examined with $\bf{I}$ oriented along various crystalline directions. From rigorous temperature-dependent MR and Hall measurements on different Hall-bar devices ($\alpha$-dependence) with various in-plane field directions ($\phi$-dependence), we uncovered three main unusual characteristics: (i) the appearance of a fourfold-symmetric component in the low-temperature $\phi$-dependent MR; (ii) rapid changes in the variations of $\phi$-dependent MR and Hall signals at low temperatures; and (iii) anomalous relative phase changes between $\phi$-dependent MR and Hall signals. To investigate these unusual characteristics, rigorous band structure calculations were performed with $\bf{M}$ oriented along various crystalline directions in the film plane. These calculations reveal significant changes in both the number and energy of Weyl nodes relative to the Fermi surface in SRO. The influence of these changes on the chiral anomaly, along with a direct comparison to our observed $\bf{I}$-direction-dependent in-plane MR and Hall behaviors, will be addressed.           

%For $\bf{I}$ along the crystalline principal axes of monoclinic [001]$_{\rm o}$ and [1\=10]$_{\rm o}$, a large negative magnetoresistance (MR) was observed when the applied field was $\bf{H} \parallel \bf{I}$ as expected for the enhanced conductivity due to the chiral anomaly \cite{SS2013,Burkov2015}. In striking contrast, when $\bf{I}$ was along a direction of about 45$^{\rm o}$ away from [001]$_{\rm o}$, we uncovered a pronounced fourfold-symmetric component in the low-temperature in-plane MR; specifically, negative MR occurred for $\bf{H}$ along [001]$_{\rm o}$ and [1\=10]$_{\rm o}$ instead. To investigate this discrepancy, rigorous band structure calculations were performed with $\bf{M}$ oriented along various crystalline directions in the film plane. These calculations reveal significant changes in both the number and energy of Weyl nodes relative to the Fermi surface in SRO. The influence of these changes on the chiral anomaly, along with a direct comparison to our observed $\bf{I}$-direction-dependent in-plane MR and Hall behaviors, will be addressed.         
                        
\section*{Results}
\subsection*{The resistivity anisotropy}
\begin{figure}[ht]
\centering
\includegraphics[width=\linewidth]{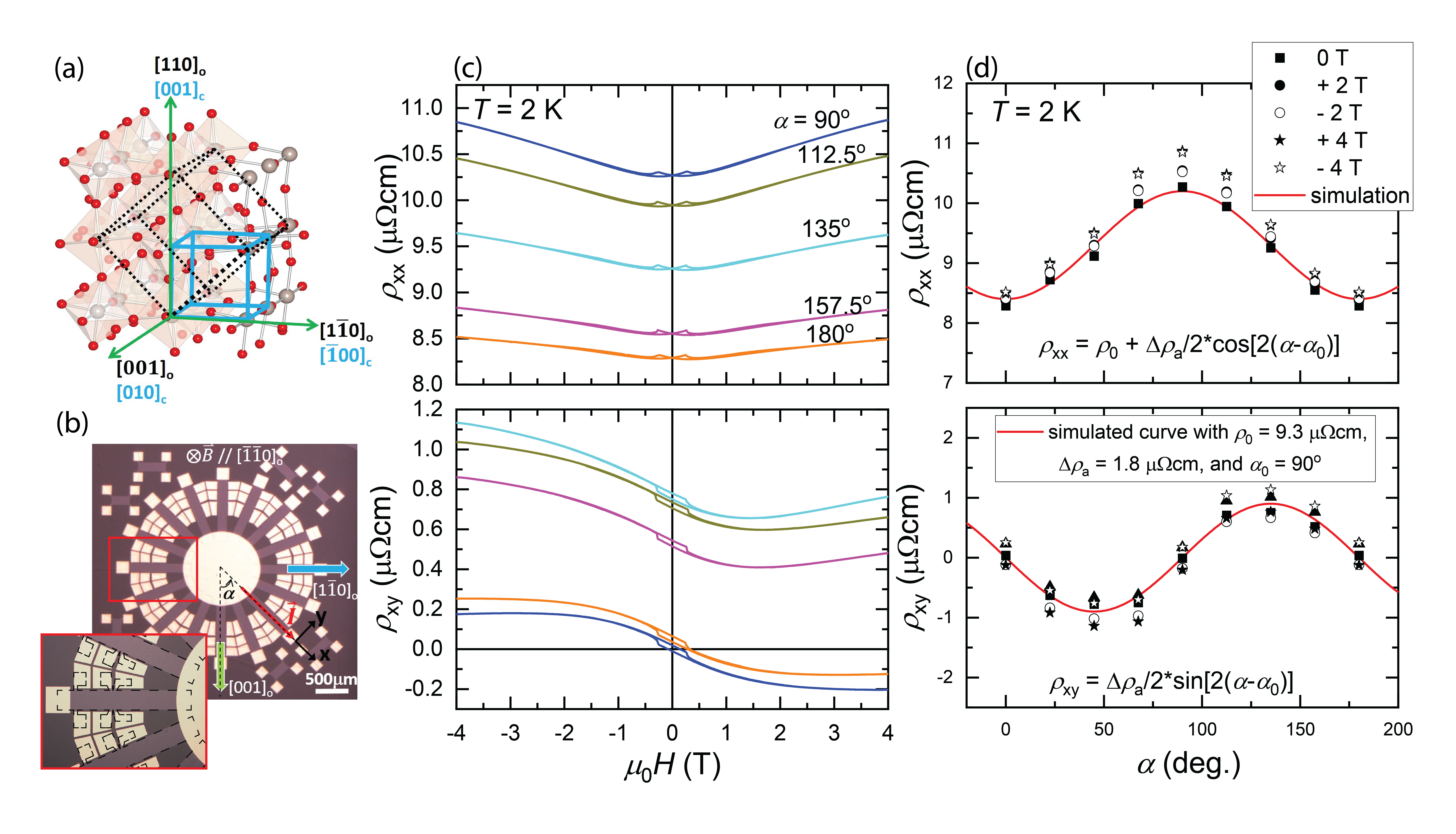}
  \caption{Resistivity anisotropy in the SRO thin film. (a) An illustration of the crystal structure of the monoclinic SRO thin film. The black dotted lines and light blue solid lines correspond to the unit cells for the monoclinic and pseudocubic structures, respectively. (b) shows an optical image of a sunbeam-shaped SRO device. The green and blue arrows indicate the two principal axes of [001]$_{\rm o}$ and [1\=10]$_{\rm o}$, respectively. The lower left inset is a blowup view of the red box, where the black dashed lines enclose the SRO Hall-bar regions after the argon-ion milling. The upper and lower panels of (c) show the field-dependent $\rho_{\rm xx}$ and $\rho_{\rm yx}$, respectively, at $T$ = 2 K, where different line colors correspond to data acquired at different $\alpha$ values. The resulting $\alpha$-dependences of $\rho_{\rm xx}$ and $\rho_{\rm yx}$ at different field strengths are plotted in the upper and lower panels of (d), respectively. Different symbols correspond to various field strengths applied along the film out-of-plane direction ([110]$_{\rm o}$), and the red lines are simulated curves based on a resistivity anisotropy model.         
  }
  \label{OFPMR}
\end{figure}

Figure \ref{OFPMR}(c) shows the field-dependent resistivity ($\rho_{\rm xx}$) and Hall resistivity ($\rho_{\rm yx}$) at $T$ = 2 K with $\mu_{\rm0}\bf{H}$ along [110]$_{\rm o}$. The subscripts x and y represent local orthogonal coordinates, with $\bf{x}$ defined along the $\bf{I}$, as indicated by the black arrows in Fig. \ref{OFPMR}(b). The $\alpha$-dependence of $\rho_{\rm xx}$ and $\rho_{\rm yx}$ can thus be extracted at various field strengths and is shown in Fig. \ref{OFPMR}(d). These data can be well fitted by a resistivity anisotropy model of $\rho_{\rm xx}$ = $\rho_0$+($\Delta\rho_{\rm a}$/2)cos[2($\alpha-\alpha_0$)] and $\rho_{\rm yx}$ = ($\Delta\rho_{\rm a}$/2)sin[2($\alpha-\alpha_0$), yielding an anisotropy magnitude of $\Delta\rho_{\rm a}/\rho_0 \approx$ 19 \% at $T$ = 2 K. We note that $\Delta\rho_{\rm a} \approx$ 1.8 $\mu\Omega$cm is nearly sixty-fold larger than the intrinsic anomalous Hall resistivity at zero field, while the MR ($[\rho(H)/\rho(0)-1]$) is positive, as demonstrated in the upper panel of Fig. \ref{OFPMR}(c). The magnetic easy axis of our SRO thin film with a thickness of about 13.7 nm was found to be oriented about 20 degrees away from the [110]$_{\rm o}$ \cite{KarPRX2024}, which may give rise to the observed resistivity anisotropy. However, as demonstrated in Fig. \ref{OFPMR}(d), the observed $\Delta\rho$ shows a weak dependence on the magnetic field strength up to 4 T along [110]$_{\rm o}$ that aligns the $\bf{M}$ more closely to [110]$_{\rm o}$. Therefore, the origin of the large resistivity anisotropy is more likely associated with the crystalline anisotropy and the electronic correlation effects \cite{Wu2020}. We also note that the observed resistivity anisotropy further supports the presence of an untwinned and dominant single-structural domain in our SRO thin films.    
\subsection*{The $\bf{I}$-orientation-dependent in-plane MR and Hall measurements}
\begin{figure}[ht]
\includegraphics[width=\linewidth]{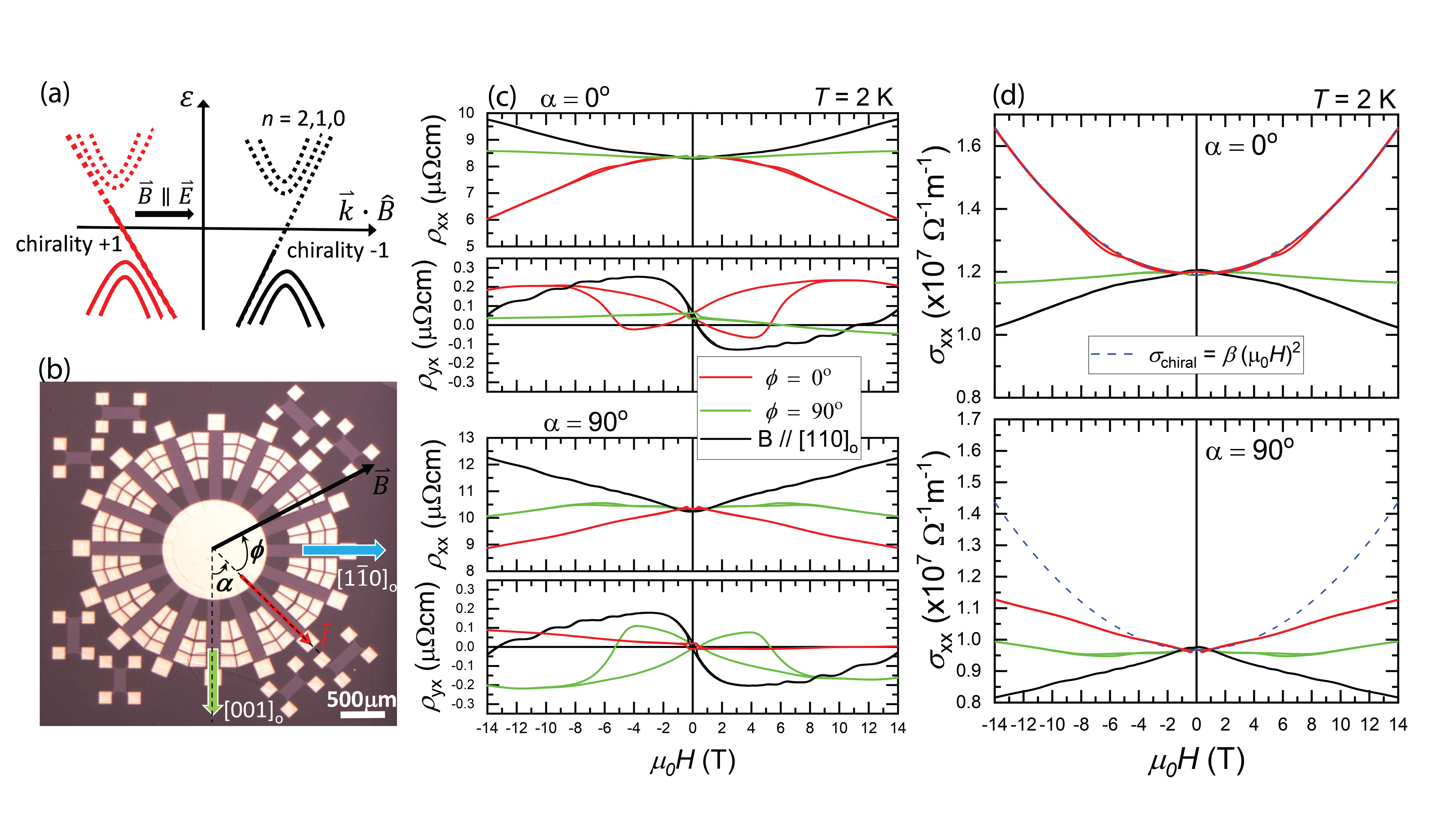}
  \caption{ In-plane MR and Hall effect in the SRO thin film at $T$ = 2 K. (a) A minimum model of a WSM and the chiral anomaly, showing non-conserving chiral charges under the condition of $\bf{B} \parallel \bf{E}$. As illustrated in (b), $\alpha$ is the angle between the $\bf{I}$ and [001]$_{\rm o}$, and $\phi$ is the angle between the in-plane $\bf{H}$ and $\bf{I}$. (c) The upper (lower) panel shows the field-dependent $\rho_{\rm xx}$ and $\rho_{\rm yx}$ for the $\alpha$ = 0$^{\rm o}$ (90$^{\rm o}$) Hall-bar device. The red and green curves correspond to data acquired with an in-plane $\bf{H}$ at $\phi$ = 0$^{\rm o}$ and 90$^{\rm o}$, respectively. The black curves are MR and Hall data with an out-of-plane $\bf{H}$ along [110]$_{\rm o}$ for comparison. The corresponding magnetoconductivity ($\sigma(H)$) for the $\alpha$ = 0$^{\rm o}$ (90$^{\rm o}$) Hall-bar device is shown in the upper (lower) panel of (d). The blue dashed lines are fitting curves calculated using the same $\beta$ parameter in the formula of $\sigma_{\rm chiral} = \beta (\mu_{\rm 0}H)^2$, representing the enhanced conductivity arising from the chiral anomaly.         
}
  \label{IPAMR}
\end{figure}

Magnetotransport measurements at $T$ = 2 K with an in-plane magnetic field were carried out and are plotted in Fig. \ref{IPAMR}, where the $\alpha$ and $\phi$ are defined as the angles from $\bf{I}$ to [001]$_{\rm o}$ and the magnetic field direction, respectively. The in-plane field dependence of $\rho_{\rm xx}$ and $\rho_{\rm yx}$ for $\alpha$ = 0$^{\rm o}$ ($\bf{I} \parallel$ [001]$_{\rm o}$) and 90$^{\rm o}$ is shown in the upper and lower panels of Fig. \ref{IPAMR}(c), respectively. A large negative longitudinal MR with $\phi$ = 0$^{\rm o}$ (NLMR) is observed for both $\alpha$ values, as expected for a topological Weyl semimetal exhibiting the chiral anomaly \cite{SS2013,Burkov2015}. As the angle $\phi$ increases from 0$^{\rm o}$ to 90$^{\rm o}$, the magnitude of the NLMR progressively decreases, which is in sharp contrast to the positive MR observed for $\bf{H}$ along the film surface normal ([110]$_{\rm o}$). On the other hand, the profiles of the field-dependent Hall resistivity $\rho_{\rm yx}$ for various $\phi$ values turn out to be dictated by the $\bf{H}$ orientation with respect to the crystalline axes. For $\bf{H} \parallel$ [001]$_{\rm o}$, $\rho_{\rm yx}$ exhibits a large symmetric component with respect to $\mu_{0}H$, indicated by the red (green) curve in the upper (lower) panel of Fig. \ref{IPAMR}(c). The hysteretic $\rho_{\rm yx}$ behavior for $\mu_{0}H \leq$ 8 T indicates an $\bf{H}$-orientation-dependent $\bf{M}$ rotation process that has been reported previously \cite{Klein2013}. The corresponding magnetoconductivity ($\sigma_{\rm xx}$) at $T$ = 2 K is plotted in Fig. \ref {IPAMR}(d), sharing the same color code as the data in Fig. \ref{IPAMR}(c). The blue dashed lines are fitting curves based on the formula of  
$\sigma_{\rm chiral} = \beta (\mu_0H)^2$ using the same $\beta \approx$ 2.4$\times 10^4$ $\Omega^{-1}$m$^{-1}$T$^{-2}$, where $\sigma_{\rm chiral}$ is the enhanced conductivity due to the chiral anomaly \cite{SS2013,Burkov2015}. For $\alpha$ = 90$^{\rm o}$, the $H^{2}$ dependence of $\sigma_{\rm chiral}$ is observed for field strengths below about 4 T, above which it practically shows a $H$-linear dependence. On the other hand, a clear $H^{2}$-dependent $\sigma_{\rm chiral}$ was found for $\alpha$ = 0${^{\rm o}}$ up to 14 T.      

\begin{figure}[ht]
\includegraphics[width=\linewidth]{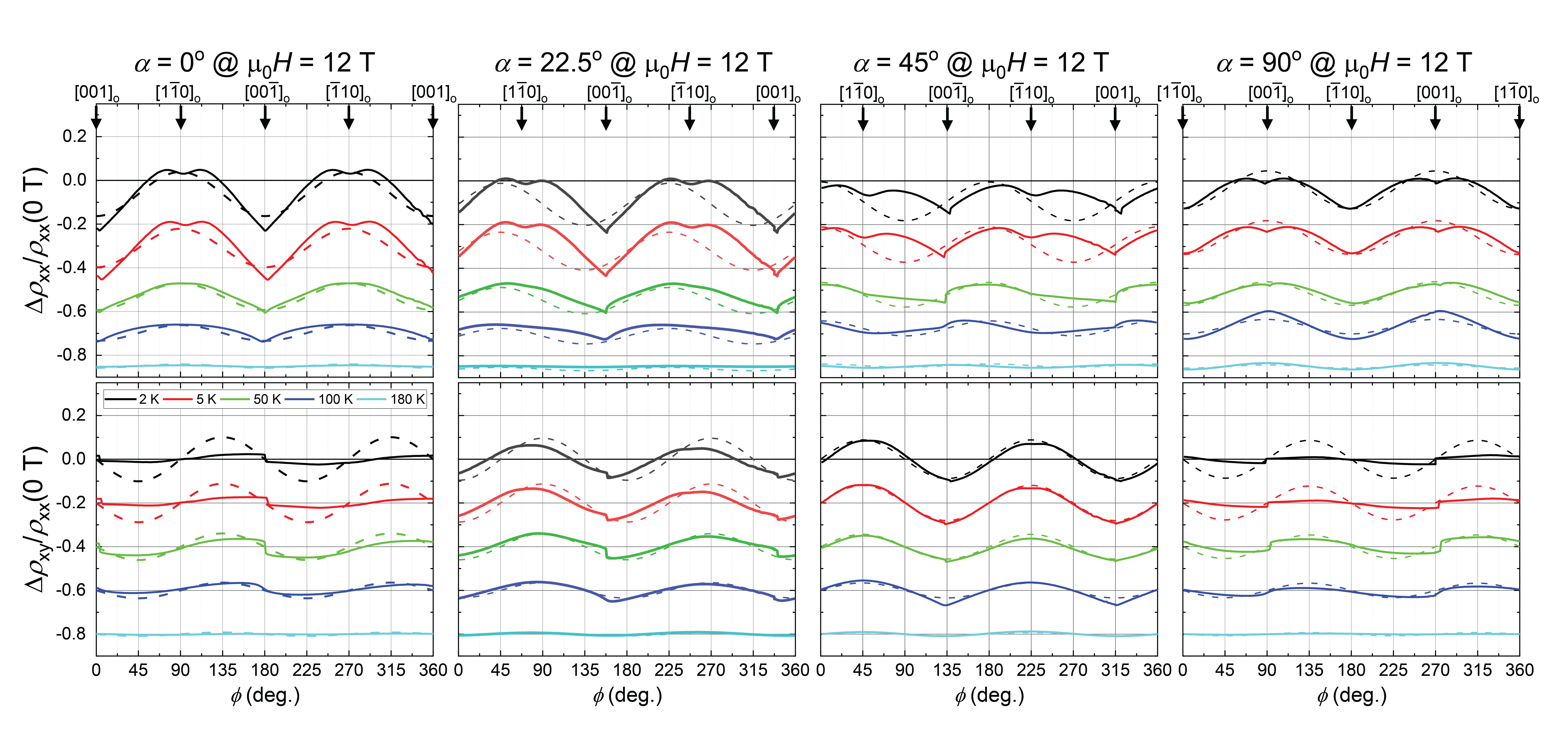}
  \caption{In-plane MR and Hall data in the SRO thin film for four Hall-bar devices. Data are shown for $\alpha$ = 0$^{\rm o}$, 22.5$^{\rm o}$, 45$^{\rm o}$, and 90$^{\rm o}$ from left to right. The upper and lower panels show the $\phi$-dependent $\Delta\rho_{\rm xx}/\rho_{\rm xx}$(0 T) and $\Delta\rho_{\rm yx}/\rho_{\rm xx}$(0 T), respectively. The solid and dashed lines are the experimental data and simulated noncrystalline AMR curves, respectively. The color code corresponds to different $T$s of 2, 5, 50, 100, and 180 K. For $T \geq$ 5 K, the data are vertically shifted for clarity. The black arrows indicate the $\phi$ values for $\bf{H}$ along the principal axes of [001]$_{\rm o}$ and [1\={1}0]$_{\rm o}$. An apparent fourfold-symmetric component appears in the $\phi$-dependent $\Delta\rho_{\rm xx}$ at low $T$s. 
  }
  \label{phidepAMR}
\end{figure}

Comprehensive $\phi$-dependent $\Delta\rho_{\rm xx}(H)/\rho_{\rm xx}$(0 T) and $\Delta\rho_{\rm yx}(H)/\rho_{\rm xx}$(0 T) at $\mu_{0}H$ = 12 T are shown in the upper and lower panels, respectively, of Fig. \ref{phidepAMR} for four $\alpha$ values of 0$^{\rm o}$, 22.5$^{\rm o}$, 45$^{\rm o}$, and 90$^{\rm o}$. Here, $\Delta\rho_{\rm xx(yx)}(H) \equiv \rho_{\rm xx(yx)}(H) - \rho_{\rm xx(yx)}$(0 T). The solid lines are experimental data at different $T$s ranging from 2 to 180 K, and the data for $T \geq$ 5 K are vertically shifted for clarity. The dashed lines are simulated in-plane MR and Hall curves based on the conventional noncrystalline anisotropic MR (AMR) effect in a magnetic system. Aside from a phase offset, the same $T$-dependent parameters were used for the simulated curves across all four different $\alpha$ values. For both $\alpha$ = 0$^{\rm o}$ and 90$^{\rm o}$, $\rho_{\rm xx}$ is largely suppressed by the applied $\bf{H}$ along the $\bf{I}$ ($\phi$ = 0$^{\rm o}$ or 180$^{\rm o}$), which is in sharp contrast to the nearly $\phi$-independent $\Delta\rho_{\rm yx}$ data shown in the lower panels of Fig. \ref{phidepAMR}. Such a dramatic difference in the $\phi$-dependent $\Delta\rho_{\rm yx}$ thus strongly suggests that our observed AMR in SRO cannot be attributed to the conventional noncrystalline AMR effect arising from $s$-$d$ scattering with spin-orbit coupling \cite{McGuire1975,AMR_Karel,Lin2021}. However, we note that the negative MR for $\alpha$ = 22.5$^{\rm o}$ and 45$^{\rm o}$ occurs when $\bf{H}$ is applied along either [001]$_{\rm o}$ or [1\=10]$_{\rm o}$, rather than at $\phi$ = 0$^{\rm o}$ or 180$^{\rm o}$ with $\bf{H} \parallel \bf{I}$, where $\Delta\rho_{\rm yx}$ also exhibits a similar $\phi$ dependence as shown in the lower panels of Fig. \ref{phidepAMR}. Such a notable fourfold-symmetric component in the $\Delta\rho_{\rm xx}/\rho_{\rm xx}$(0 T) appears at low $T$s for all $\alpha$ values (main characteristic (i)), and the four local minima in $\Delta\rho_{\rm xx}/\rho_{\rm xx}$(0 T) occur at the $\phi$ values corresponding to $\bf{H}$ along the principal axes of [001]$_{\rm o}$ and [1\={1}0]$_{\rm o}$, indicated by the black arrows in the upper panels of Fig. \ref{phidepAMR}.                               
\subsection*{The temperature-dependent measurements}
\begin{figure}[ht]
\includegraphics[width=\linewidth]{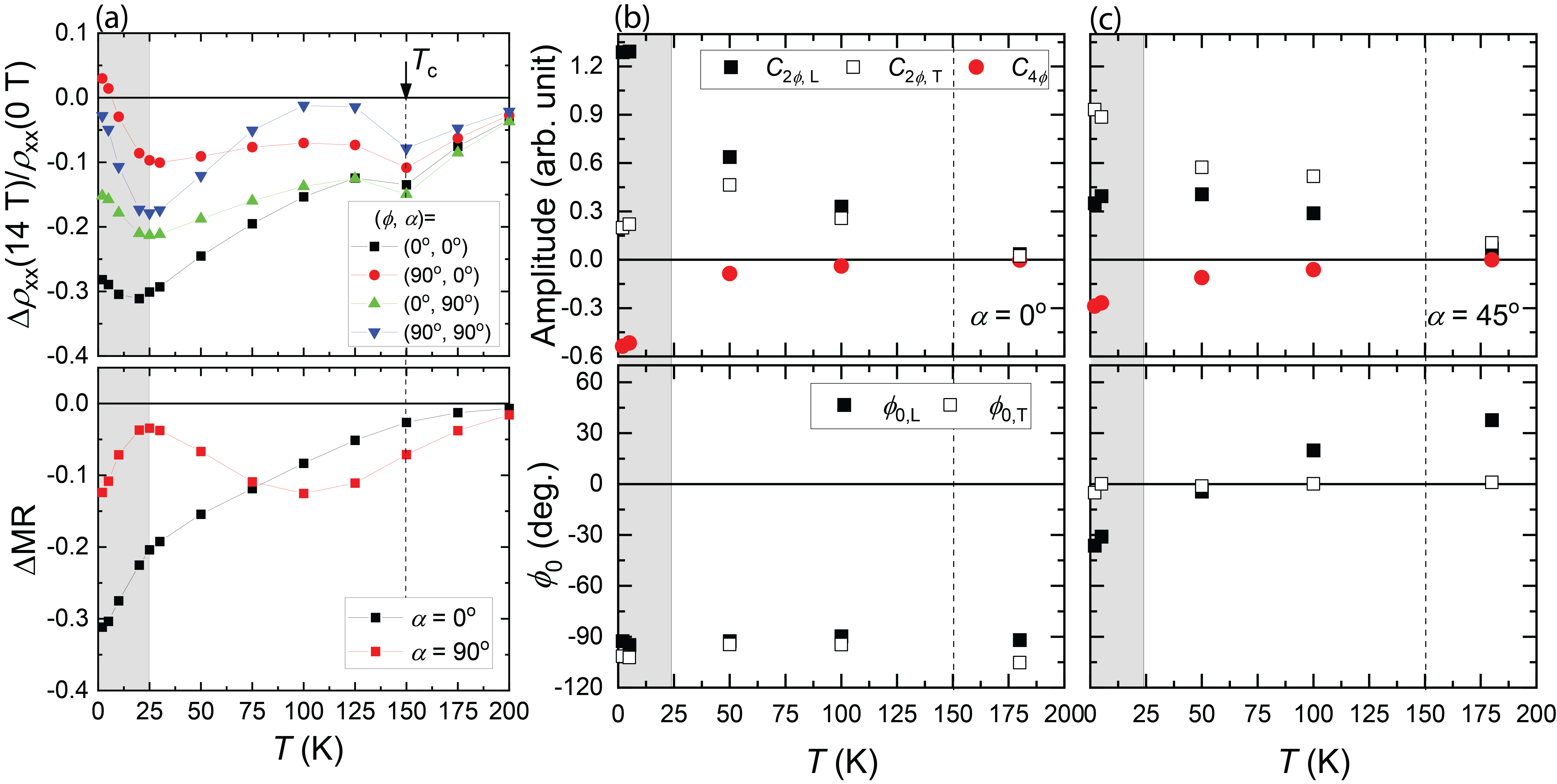}
  \caption{$T$-dependent in-plane MR and extracted AMR parameters in the SRO thin film. (a) The upper panel shows the experimental MR at 14 T ($\Delta\rho_{\rm xx}$(14 T)/$\rho_{\rm xx}$(0 T)) for four different sets of ($\phi$, $\alpha$). For $T \leq$ 25 K, indicated by a gray-shaded regime, rapid drops in the MR were observed for $\phi$ = 90$^{\rm o}$. The corresponding $\Delta$MR for $\alpha$ = 0$^{\rm o}$ and 90$^{\rm o}$ is shown in the lower panel of (a), indicating a growing NLMR at low temperatures that is consistent with the enhanced conductivity due to the chiral anomaly in a WSM. (b) and (c) show the $T$-dependent AMR parameters ($C_{2\phi,\rm L}$, $C_{2\phi,\rm T}$, $C_{4\phi}$, $\phi_{0,\rm L}$, and $\phi_{0,\rm T}$) for $\alpha$ = 0$^{\rm o}$ and 45$^{\rm o}$, respectively, based on the phenomenological AMR formula, where the subscript L(T) refers to the in-plane MR(Hall) signals. For all $\alpha$ values, a sizable $C_{4\phi}$ value appears only in the in-plane MR at low $T$s. An unusual $T$-dependent phase difference between $\phi_{0,\rm L}$ and $\phi_{0,\rm T}$ is observed for $\alpha$ = 45$^{\rm o}$.    
  }
  \label{TdepLMR}
\end{figure}

The temperature dependence of the in-plane MR at $\mu_{0}H$ = 14 T ($\Delta\rho_{\rm xx}$(14 T)/$\rho_{\rm xx}$(0 T)) for four representative ($\phi$, $\alpha$) values is shown in the upper panel of Fig. \ref{TdepLMR}(a). The red solid spheres and blue down-triangles represent the MR data for $\phi$ = 90$^{\rm o}$ ($\bf{H} \perp \bf{I}$) with $\alpha$ = 0$^{\rm o}$ and 90$^{\rm o}$, respectively, where a rapid drop in the magnitude of the negative MR is observed for $T \leq$ 25 K. For ($\phi$, $\alpha$) = (90$^{\rm o}$,0$^{\rm o}$), the MR changes to a small positive value for $T \leq$ 6 K. In contrast, a large NLMR persists down to the lowest measured temperature of about $T \approx$ 1.5 K for $\alpha$ = 0$^{\rm o}$ and 90$^{\rm o}$ shown as black solid squares and green up-triangles, respectively. The resulting $\Delta$MR $\equiv$ MR($\phi$ = 0$^{\rm o}$)$-$MR($\phi$ = 90$^{\rm o}$) at $\mu_{0}H$ = 14 T for $\alpha$ = 0$^{\rm o}$ and 90$^{\rm o}$ is shown in the lower panel of Fig. \ref{TdepLMR}(a). Although $\Delta$MR remains negative across the entire temperature range up to 200 K, its rapid increase below 25 K (indicated by the gray-shaded region in Fig. \ref{TdepLMR}) marks a distinct regime, and this behavior differs from the high temperature regime in which conventional spin-dependent scattering and spin-fluctuations dominate. Furthermore, the NLMR effect observed for $T \leq$ 25 K shows no apparent dependence on the location of contact electrodes, thereby excluding the current jetting effect \cite{Liang2018chiral,Ong2021} (see Supplementary Note 1). Additionally, the corresponding change in the magnetoconductivity (Fig. \ref{IPAMR} (d)) is more than two orders of magnitude larger than $e^2$/$h$, suggesting that our observed NLMR does not fall within the weak localization regime as is typically expected in disordered metals \cite{Maekawa1981}.       
\subsection*{The AMR model and fitting}
In general, the resistivity tensor can be expanded in a power series of the directional cosines of $\bf{M}$, containing crystalline contributions in accord with the crystalline symmetry \cite{Ding2023}. The AMR can thus be described by $\Delta\rho_{\rm xx}/\rho_{\rm xx}$(0T) = $C_{2\phi,\rm L}$ cos2($\phi + \phi_{0,\rm L}$) + $C_{4\phi}$ cos4$\phi$ and $\Delta\rho_{\rm yx}/\rho_{\rm xx}$(0 T) = $C_{2\phi,\rm T}$ sin2($\phi + \phi_{0,\rm T}$), where the subscripts L and T refer to the in-plane MR and Hall resistivity, respectively. The twofold-symmetric component $C_{2\phi,\rm L(T)}$ derives from the noncrystalline and uniaxial crystalline AMR contributions, while $C_{4\phi}$ relates to the cubic crystalline AMR \cite{AMR_Karel}. This reduces to the noncrystalline AMR shown as dashed lines in Fig. \ref{phidepAMR} by setting $C_{4\phi}$ = 0, $C_{2\phi,\rm L}$ = $C_{2\phi,\rm T}$ and $\phi_{0,\rm L}$ = $\phi_{0,\rm T}$. The extracted $T$-dependent AMR parameters are shown in Figs. \ref{TdepLMR}(b) and (c) for $\alpha$ = 0$^{\rm o}$ and 45$^{\rm o}$, respectively (see Supplementary Note 3). For $T \geq$ 25 K, the observed $\phi$-dependent $\Delta\rho_{\rm xx}$ and $\Delta\rho_{\rm yx}$ essentially follow the phenomenological equations with a ratio of $C_{2\phi,\rm T}/C_{2\phi,\rm L} \approx$ 1 and a vanishing $C_{4\phi}$ for all $\alpha$ values, which is consistent with the monoclinic crystalline symmetry in the (110)-oriented SRO thin film below the Curie temperature ($T_{\rm c} \approx$ 150 K) \cite{KarPRX2024}. However, the ratio of $C_{2\phi,\rm T}/C_{2\phi,\rm L}$ changes rapidly with decreasing $T$ to about 0.2 and 2.3 at $T \leq$ 25 K for $\alpha$ = 0$^{\rm o}$ and 45$^{\rm o}$, respectively (main characteristic (ii)). We also note that, for $\alpha$ = 45$^{\rm o}$, the corresponding $\phi_{0,\rm L}$ shows a strong $T$-dependence, giving rise to an unusual phase difference between the in-plane MR and Hall resistivity, as shown in the lower panel of Fig. \ref{TdepLMR}(c) (main characteristic (iii)). Surprisingly, the fourfold-symmetric component $C_{4\phi}$ appears for all $\alpha$ values and increases rapidly for $T \leq$ 25 K, which is incompatible with the monoclinic crystalline symmetry of the SRO thin film. Similar behavior was previously reported in Heusler alloys \cite{Oogane2018} and perovskite nitrides \cite{Tsunoda2010}, and the emergence of the $C_{4\phi}$ contribution at low $T$s was attributed to the structural distortion and changes in $d$-orbital occupancy, which is unlikely applicable to the monoclinic SRO. In addition, for $T \leq$ 25 K, we also note that the magnitudes of $C_{4\phi}$ and $C_{2\phi,\rm L}$ are comparable for $\alpha$ = 45 $^{\rm o}$, which is in sharp contrast to the overwhelming $C_{2\phi,\rm L}$ contribution observed for $\alpha$ = 0$^{\rm o}$ and 90$^{\rm o}$. 
%As revealed by our band calculations shown in Fig. \ref{bandcal}, the interplay between the chiral anomaly and tunable Weyl nodes via the control of $\bf{M}$-orientation in the SRO thin film is responsible for the observed $C_{4\phi}$ contribution at low $T$s, as well as the unusual amplitude and phase differences between the $\phi$-dependent in-plane MR and Hall signals. 
     
\section*{Discussion}
\begin{figure}[ht]
\includegraphics[width=\linewidth]{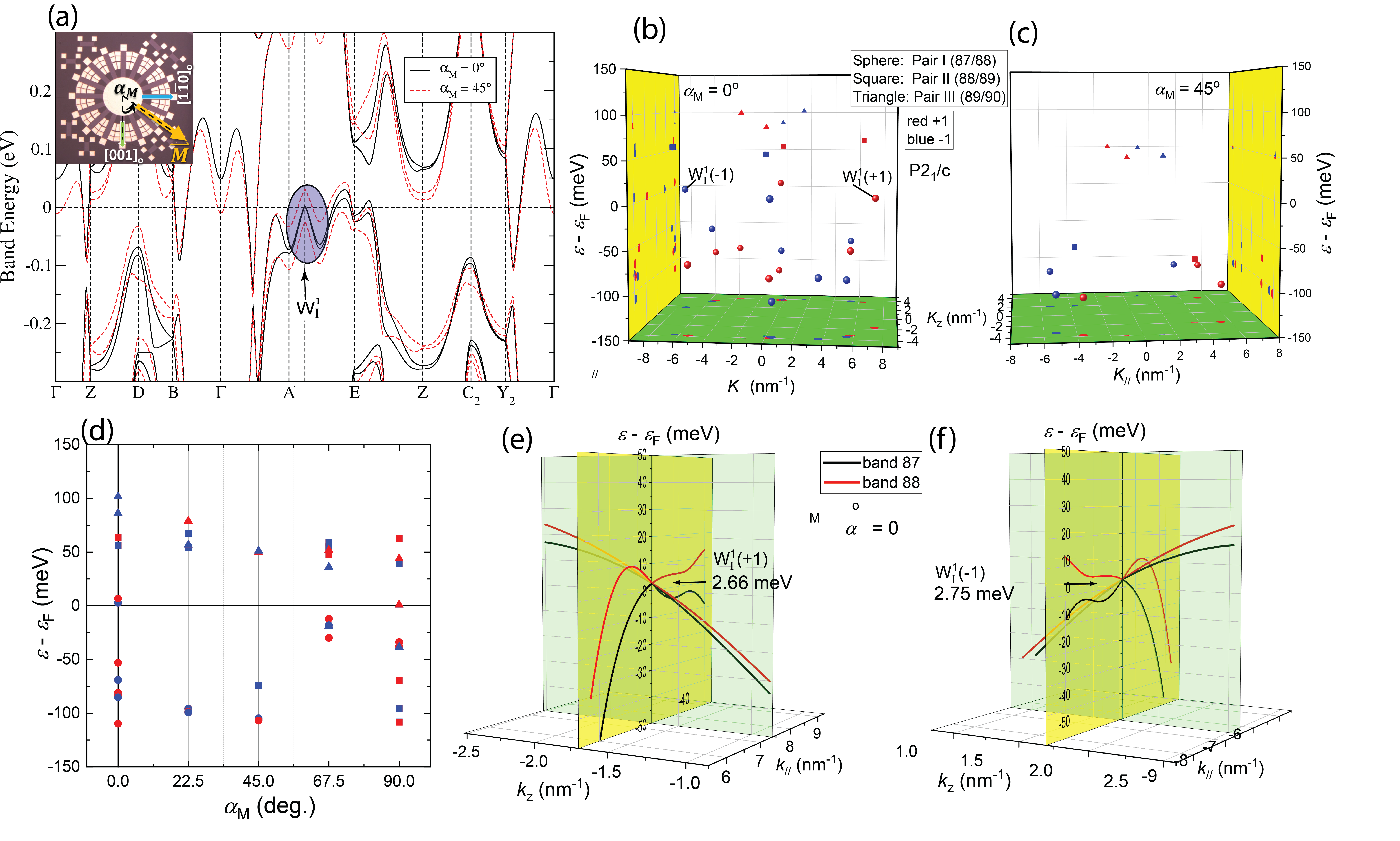}
\caption{Calculated Weyl-node distribution for various $\bf{M}$ orientations. (a) The black solid and red dashed lines are the calculated electronic band structures for $\alpha_{\rm M}$ = 0$^{\rm o}$  and 45$^{\rm o}$, respectively. The angle $\alpha_{\rm M}$ is defined as the angle between $\bf{M}$ and [001]$_{\rm o}$ as illustrated in the upper left inset. The calculated Weyl-node locations for $\alpha_{\rm M}$ = 0$^{\rm o}$ and 45$^{\rm o}$ are shown in (b) and (c), respectively. The different symbols correspond to Weyl nodes from different pairs of bands, and the symbol colors of red and blue represent the chiral charges of +1 and -1, respectively. The $W_{\rm I}^1$($\pm$1) pair is located within the blue shaded region in (a), which is the closest Weyl-node pair to the Fermi surface for $\alpha_{\rm M}$ = 0$^{\rm o}$. (d) plots the Weyl-node energy ($\varepsilon-\varepsilon_{\rm F}$) versus $\alpha_{\rm M}$. The corresponding band dispersions for $W_{\rm I}^1$($\pm$1) projected on two orthogonal planes cutting across the Weyl nodes are shown in (e) and (f).          
}
\label{bandcal}
\end{figure}

The magnetic easy axis for our 13.7 nm SRO thin film is about 20$^{\rm o}$ away from [110]$_{\rm o}$, tilting toward [\={1}10]$_{\rm o}$ \cite{Klein1996,Koster2012,Ziese2010,Klein2013,KarPRX2024}. We estimated the anisotropy field to be about 10 T at $T$ = 5 K (see Supplementary Note 2); it is thus reasonable to assume that $\bf{M}$ is forced to align mostly along the field direction for $\mu_{0}H \geq$ 10 T. We performed rigorous band structure calculations with $\bf{M}$ lying within the (110)$_{\rm o}$ plane, where $\alpha_{\rm M}$ is defined as the angle between $\bf{M}$ and [001]$_{\rm o}$ as illustrated in the upper inset of Fig. \ref{bandcal}(a). The calculated band structures for $\alpha_{\rm M}$ = 0$^{\rm o}$ and 45$^{\rm o}$ are shown as black solid and red dashed lines, respectively, in Fig. \ref{bandcal}(a), revealing notable variations in the band structures with the $\bf{M}$ orientation. As highlighted in the blue-shaded region of Fig. \ref{bandcal}(a), the $W_{\rm I}^{1}$ Weyl-node, closest to the Fermi surface for $\alpha_{\rm M}$ = 0$^{\rm o}$, is found to split apart by about 50 meV from the Fermi surface when $\alpha_{\rm M}$ is changed to 45$^{\rm o}$. Such dramatic $\alpha_{\rm M}$-dependent band structures in SRO are further shown in Fig. \ref{bandcal}(b) and (c) for $\alpha_{\rm M}$ = 0$^{\rm o}$ and 45$^{\rm o}$, respectively, where the calculated Weyl-node distributions within an energy window of $\pm$ 100 meV near the Fermi surface are included. One notable difference is the smaller number of Weyl nodes near the Fermi surface when $\bf{M}$ points away from the principal crystalline axes. In addition, the Weyl node closest to the Fermi surface for $\alpha_{\rm M}$ = 45$^{\rm o}$ is located at about $\varepsilon-\varepsilon_{\rm F}$ = 50 meV, which is significantly larger than the value for $\alpha_{\rm M}$ = 0$^{\rm o}$ ($\varepsilon-\varepsilon_{\rm F}$ = 2.66 meV, labeled as $W_{\rm I}^{1}(+1)$ in Fig. \ref{bandcal}(b)). The detailed $\alpha_{\rm M}$ dependence of the Weyl-node locations is shown in Fig. \ref{bandcal}(d), where an apparent lack of Weyl nodes near the Fermi surface was found for $\alpha_{\rm M}$ near 45$^{\rm o}$. These results echo a number of previous reports on topological band structure tuning via the control of $\bf{M}$ magnitude and orientation, which is an intriguing property in magnetic topological systems \cite{BFB2022,Fang2003,Yang2020}. For $\alpha_{\rm M}$ = 0$^{\rm o}$, the calculated band dispersions near the Weyl-node pair of $W_{\rm I}^{1}(+1)$ and $W_{\rm I}^{1}(-1)$ are shown in Fig. \ref{bandcal}(e) and (f), respectively, where the red and black lines are the projected bands on two orthogonal planes cutting across the Weyl nodes. These results support the presence of a Weyl-node pair tilting along the $k_{\rm z}$ direction.
                 
Considering a minimum WSM model as illustrated in Fig. \ref{IPAMR}(a), the enhanced conductivity due to an imbalanced total chiral charge requires the conditions of $\tau_{\rm inter} \gg \tau_{\rm intra}$ and $\bf{H} \parallel \bf{I}$, where $\tau_{\rm inter}$ ($\tau_{\rm intra}$) refers to the intervalley (intravalley) scattering lifetime \cite{SS2013}. In the weak field regime, this can be explicitly described by $\sigma_{\rm chiral} = \beta (\mu_0H)^2$, where $\beta = \frac{e^4\upsilon_{\rm F}^3\tau_{\rm inter}}{2\pi h (\varepsilon - \varepsilon_{\rm F})^2}$ \cite{SS2013,Burkov2015} and $\upsilon_{\rm F}$ is the Fermi velocity. By using the values $\varepsilon_{\rm F}$ = 1.34 $\times$ 10$^5$ ms$^{-1}$ \cite{Kar2023} and ($\varepsilon - \varepsilon_{\rm F}$) = 2.66 meV from our band structure calculations, the extracted $\tau_{\rm inter}$ gives a lower bound of $\approx$ 12 ps which is more than 30 times larger than the quantum lifetime extracted from the quantum oscillation experiments at $T$ = 2 K \cite{Kar2023}, thereby justifying the criterion of $\tau_{\rm inter} \gg \tau_{\rm intra}$ for the occurrence of the chiral anomaly. The observed large NLMR for $\alpha$ = 0$^{\rm o}$ and 90$^{\rm o}$ is thus expected due to the chiral anomaly and is consistent with several earlier works \cite{Rao1998,Taki2020}. On the other hand, the nearly twenty-fold increase in the ($\varepsilon-\varepsilon_{\rm F}$) value for the closest Weyl node at $\alpha_{\rm M}$ = 45$^{\rm o}$ largely suppresses the $\sigma_{\rm chiral}$ contribution, which qualitatively accounts for the absence of NLMR at $\alpha$ = 45$^{\rm o}$ shown in Fig. \ref{phidepAMR}. Remarkably, the negative MR behavior for $\alpha$ = 45$^{\rm o}$ was found to emerge when $\bf{M}$ was reoriented by $\bf{H}$ along either [001]$_{\rm o}$ or [1\={1}0]$_{\rm o}$ (black arrows in Fig. \ref{phidepAMR}), where the criteria of having a Weyl node close to the Fermi surface and a finite $\bf{H}$ component along $\bf{I}$ can both be satisfied for the occurrence of the chiral anomaly. The $\alpha$-dependent in-plane MR can thus be qualitatively understood within a unified picture, and the NLMR due to the chiral anomaly can be effectively tuned by shifting the Weyl-node location via the control of the $\bf{M}$ orientation. We also remark that all MR and Hall signals were recorded and plotted using the same full-loop, field-sweeping sequence up to a 14 T field strength to ensure a consistent magnetic domain structure in the SRO thin films for comparison. In addition, the observed $\rho_{\rm xx}$ variations in resistivity anisotropy (Fig. \ref{OFPMR}(d)) and the $\phi$-dependent $\rho_{\rm xx}$ (Fig. \ref{phidepAMR}) in our SRO thin film are on the order of $\approx$ 2 $\mu\Omega$cm, which is more than one order of magnitude larger than the previously reported domain wall resistivity contribution in SRO thin films \cite{Klein2000,Feigenson2003}. We thus conclude that the domain wall resistivity contribution is negligible. 

%It has been reported previously in several other topological systems that a large planar Hall effect may arise due to Berry curvatures from Weyl nodes near the Fermi surface or chiral anomaly effect \cite{Burkov2017,Nandy2017,Liang2018}, which is not compatible with our observation of vanishing and nearly $\phi$-independent $\rho_{\rm yx}$ with in-plane field up to 14 T for $\alpha$ = 0$^{\rm o}$ and 90$^{\rm o}$ (Fig. \ref{phidepAMR}). 

Combining the resistivity anisotropy effects, the total electric field induced by a bias current density $\bf{j}$ can be expressed as $\bf{E}$ = $\Delta\rho_{\rm M}$(\^{M}$\cdot\bf{j}$)\^{M} + $\Delta\rho_{\rm a}$(\^{a}$\cdot\bf{j}$)\^{a}, where $\Delta\rho_{\rm M}$ is the resistivity difference between $\bf{j}$ parallel and perpendicular to \^{M}, and \^{a} is a unit vector along the [001]$_{\rm o}$ as deduced from the resistivity anisotropy data shown in Fig. \ref{OFPMR}(d). As demonstrated in Figs. \ref{OFPMR}(c) and (d), the anisotropy term of $\Delta\rho_{\rm a}$ dominates for $T \leq$ 25 K; that is, $\bf{E} \approx \bf{E}_{\rm a}$ = $\Delta\rho_{\rm a}$(\^{a}$\cdot\bf{j}$)\^{a}. Such a built-in anisotropic electric field $\bf{E_{\rm a}}$ along [001]$_{\rm o}$ dictates the $\phi$-dependent $\rho_{\rm yx}$ in the SRO thin film and may be related to the tilted nature of the Weyl nodes revealed by the band structure calculations shown in Figs. \ref{bandcal}(e-f) (see Supplementary Note 4). For $\alpha$ = 0$^{\rm o}$ and 90$^{\rm o}$ at low $T$s, $\rho_{\rm yx}$ is vanishingly small and shows a minor correlation with the in-plane field strength and orientation. However, as $\bf{j}$ tilts away from \^{a}, $\rho_{\rm yx}$ attains a finite value with a magnitude proportional to $|\bf{E}_{\rm a}|$sin$\alpha$, and its $\phi$ dependence appears to follow a profile similar to $\rho_{\rm xx}$ as shown in Fig. \ref{phidepAMR} for $\alpha$ = 22.5$^{\rm o}$ and 45$^{\rm o}$. The corresponding phase difference between $\phi_{0,\rm T}$ and $\phi_{0,\rm L}$ for $T \leq$ 25 K thus approaches $\pi/4$ as shown in the lower panel of Fig. \ref{TdepLMR}(c).          
The rapid increase in the $\Delta$MR below 25 K, as shown in Fig. \ref{TdepLMR}, indicates the entry into the chiral-anomaly-dominant regime, which marks a clear distinction from the conventional AMR regime at higher $T$s. This is also in line with the Fermi liquid behavior where $\rho_{\rm xx} \propto T^2$ (see Supplementary Note 5), suggesting that suppressed spin fluctuations and electron correlation in SRO may play a role \cite{Georges2013}. As revealed in several earlier works \cite{Taki2020,Kar2023,KarPRX2024}, we remark that the topological surface states may also play an important role in the charge transport of the SRO thin film with a thickness of about 10 nm at low $T$s. Our observed unusual in-plane MR and Hall behaviors in SRO at low $T$s may thus be a consequence of the combined effects of electronic correlation and topological properties, which calls for further investigation.                                 

In summary, we rigorously investigated the $\bf{I}$-direction-dependent in-plane MR and Hall effects in a sunbeam-shaped device fabricated from an untwinned ferromagnetic Weyl metal SRO thin film with exceptionally high RRR and low RR. This setup provides a strict test of the enhanced conductivity arising from the chiral anomaly under the condition of $\bf{H} \parallel \bf{I}$. For an in-plane $\bf{H}$ larger than the anisotropy field, a large NLMR is observed for $\bf{I}$-directions along the principal axes of either [001]$_{\rm o}$ ($\alpha$ = 0$^{\rm o}$) or [1\={1}0]$_{\rm o}$ ($\alpha$ = 90$^{\rm o}$), where the corresponding in-plane Hall resistivity remains vanishingly small. Such behavior is consistent with the chiral anomaly in a WSM but incompatible with the conventional AMR effect in a magnetic system. Surprisingly, such an enhanced conductivity appears under the condition of either $\bf{H} \parallel$ [001]$_{\rm o}$ or $\bf{H} \parallel$ [1\={1}0]$_{\rm o}$, resulting in the emergence of an unusual fourfold-symmetric component in the $\phi$-dependent in-plane MR at low $T$s. Furthermore, rapid variations of the $C_{2\phi,\rm T}/C_{2\phi,\rm L}$ ratio at low $T$s and anomalous $T$-dependent phase differences between $\phi_{0,\rm L}$ and $\phi_{0,\rm T}$ were uncovered. These unusual characteristics could be partly explained by our comprehensive band structure calculations with $\bf{M}$ along different crystalline directions, which indicate dramatic shifts of the Weyl nodes farther from the Fermi surface when $\bf{M}$ points away from either [001]$_{\rm o}$ or [1\={1}0]$_{\rm o}$. Nevertheless, the influence of electronic correlation and topological surface states cannot be excluded, and further studies are keenly required. 

%Nevertheless, Together with the $T$-dependent resistivity anisotropy data and tilted Weyl nodes from calculated band dispersion, a built-in anisotropic electric field $\bf{E_{\rm a}}$ and electronic correlation are both present and may play a role the unusual in-plane MR and Hall behavior in SRO thin film.            

 %we uncovered an intriguing interplay between the $\bf{M}$ orientation and Weyl-node distribution, 

\section*{Methods}
\subsection*{The SRO film growth and device fabrication}
By using the adsorption-controlled growth technique with oxide molecular beam epitaxy, SRO thin films can be grown with a low ruthenium-vacancy defect level, and the as-grown SRO thin film with a thickness of about 13.7 nm on a SrTiO$_3$ (001) substrate shows an exceptionally high RRR of $\approx$ 24 and a low RR of about 8.3 $\mu\Omega$cm at $T$ = 2 K with nearly single-structural domain as illustrated in Fig. \ref{OFPMR}(a) \cite{Nair2018,Kar2021,Kar2023}, where dashed black lines and solid blue lines indicate the unit cell of the monoclinic (distorted-orthorhombic) structure and pseudocubic structure, respectively. We note that such a high RRR and a low RR in our SRO thin film support the presence of a low level of Ru-vacancy defects, where these Ru vacancies are known to largely affect the magnetotransport and magnetic properties of SRO thin films \cite{cap2002,KT2023}. The resulting SRO thin film is (110)-oriented on the STO (001) substrate \cite{Kar2021} with the principal axis of [110]$_{\rm o}$ nearly normal to the film surface. Here, the subscripts o and c refer to the monoclinic and pseudocubic structures, respectively. The film was then patterned into a sunbeam-shaped device using standard photolithography and argon-ion milling, as shown in Fig. \ref{OFPMR}(b). The resulting sunbeam-shaped device comprises 16 Hall-bar devices with an identical geometry of 290 $\mu$m in length and 150 $\mu$m in width, enabling precise determinations of the resistivity and Hall resistivity. The $\bf{I}$-directions for adjacent Hall-bar devices differ by 22.5 degrees, and the definition of $\alpha$ is illustrated in Fig. \ref{OFPMR}(b). Au (35 nm)/Ti (10 nm) electrodes were fabricated subsequently to form contact pads. The inset of Fig. \ref{OFPMR}(b) is a blowup view of the red box, where the black dashed lines enclose the SRO thin film regions after the fabrication. 
The magnetoconductivity measurements were carried out using a rotating probe in a superconducting magnet cryostat covering a temperature range from 1.5 to 300 K and a magnetic field strength up to 14 T. In addition, the sample stage can be rotated to fulfill the condition of $\bf{H} \parallel \bf{I}$, required for the chiral anomaly, for each Hall-bar in the sunbeam-shaped device. Such an experimental setup thus enables rigorous angle-resolved measurements of anisotropic MR and the planar Hall effect in ferromagnetic Weyl metal SRO, where dependences on both $\bf{I}$ and $\bf{H}$ orientations can be determined.      
\subsection*{The theoretical electronic band structure calculations}
The electronic structure calculations for the monoclinic structure of SrRuO$_3$ were performed using the projector augmented-wave method \cite{Kresse}, as implemented in the Vienna ab-initio simulation package (VASP)
[38], within the generalized gradient approximation scheme \cite{PBE}. A 9$\times$9$\times$6 Gamma centered k-point mesh was used, together with a cutoff energy of 500 eV. The total energy convergence criterion was
10$^{-6}$ eV. The spin-orbit coupling was included in the self-consistent calculations. In all the cases, we have used direction constrained magnetic moment calculations. The effect of the on-site electronic correlations in the Ru $d$ states (4$d^4$ for Ru$^{4+}$) was taken into account by using the rotationally invariant GGA + U scheme \cite{Liechtenstein} with $U$ =3.0 eV and $J$ =0.6 eV. We have used Ru $d$-orbitals and O $p$-orbitals to
construct the Wannier functions \cite{Marzari,Mostofi} with VASP2WANNIER90 \cite{Franchini} interface. Finally, we used WannierTools \cite{QuanSheng} to search for Weyl points and to identify the chirality of each point.

\section*{Data Availability}
All the supporting data are included in the main text and also in the supplementary information. The raw data and other related data for this paper can be requested from W.L.L. (wlee@phys.sinica.edu.tw). 
\section*{Code Availability}
The input files for DFT calculations using WIEN2k are available upon request. 
\section*{Acknowledgements}
This work was supported by National Science and Technology Council of Taiwan (NSTC Grant No. 110-2112-M-002-030-MY3, 111-2112-M-001-056-MY3 and 113-2123-M-002-015) and Academia Sinica Grand Challenge Program Seed Grant (AS-GCS-113-M08).
\section*{Author Contributions}
U.K., A.K.S., E.C.H.L., W.A., F.E.C., and W.L.L. carried out the low-temperature magneto-transport measurements and data analyses. U.K. and A.K.S. grew the epitaxial SRO films. A.K.S., S.Y., C.Y.L., and C.H.H. performed the high-precision X-ray measurements at NSRRC in Taiwan. P.V.S.R. and G.Y.G. performed SRO band structure calculations. G.Y.G. and W.L.L. designed the experiment and wrote the manuscript.  
\section*{Competing interests} The authors declare no competing financial or non-financial interests.
\section*{Additional Information}
\textbf{Supplementary Information} accompanies the paper on the XXXX website (https://XXXXX).

\end{document}